\documentclass[prl,twocolumn]{revtex4}
\usepackage{amsmath,amssymb}
\usepackage{amsthm}
\usepackage{epsfig}
\usepackage{xcolor}
\usepackage{fixltx2e}
\usepackage[normalem]{ulem}
\usepackage{siunitx}

\usepackage{xr}
\graphicspath{{eps/}}            

\begin{document}

\title{Theory of modulation instability for coupled microresonators}

\author{Alena Kolesnikova$^{1}$}
\email[Corresponding author : ]{a.kolesnikova@skoltech.ru}
\author{Andrey Gelash$^{1}$}
\email[Corresponding author : ]{a.gelash@skoltech.ru}

\affiliation{$^{1}$Center of Engineering Physics, Skolkovo Institute of Science and Technology, Moscow, Russia}

\begin{abstract}
Thanks to hybridization phenomena, coupled microresonator systems enable modulation instability (MI) of the continuous wave background field across all possible dispersion combinations: anomalous-anomalous, normal-normal, and anomalous-normal microrings. Here, using the Lugiato-Lefever model in the supermode basis for two coupled microresonators with arbitrary second-order dispersion coefficients, arbitrary losses, and coupling strength, we derive a general eigenvalue problem for the $4\times 4$ matrix. We consider the regime of resonance splitting, in which the pump couples only to one of the two supermodes, allowing us to obtain analytical expressions for the stationary continuous-wave field distributed over the coupled system. When the microresonators have similar geometry, the antidiagonal elements of the eigenvalue-problem matrix nearly vanish, and we obtain a fully analytical solution for the MI gain rate for the entire range of coupled microresonator devices. Finally, we investigate cases in which the analytical treatment must be supplemented with more accurate numerical solutions of the MI eigenvalue problem, though even then the analytical theory captures the main features of the four-wave mixing process and the gain-rate behavior.
\end{abstract}
\maketitle

\section{Introduction}
Modulation instability (MI) is a ubiquitous nonlinear wave process which, in its simplest manifestation, consists of the interaction between a high-power carrier frequency $\omega_p$ and a pair of symmetric sidebands at frequencies $\omega^{\pm} _{p}= \omega_p \pm \Omega$; see e.g.~\cite{zakharov2009modulation}. Typically, MI is considered for a constant-amplitude carrier-wave (CW) background perturbed by broadband noise that interacts with the carrier at a spectrally dependent rate and grows exponentially. The degenerate four-wave mixing process $2\omega_p \rightarrow \omega^{-}_{p}+\omega^{+}_{p}$ underlying MI was theoretically described in~\cite{Lighthill1965,Whitham1965,Zakharov1968}, and the first experimental observations were reported in hydrodynamics, optics, and electrical lines~\cite{Benjamin1967,Bespalov1966,zagryadskaya1968}. In conservative nonlinear systems such as described by the focusing nonlinear Schr{\"o}dinger equation, MI can be triggered on any CW background~\cite{Zakharov1968}. At the same time, in the presence of dissipation, the MI becomes a threshold process bounded by a minimum required CW power. In driven-dissipative systems, long-term MI often results in the formation of stationary patterns such as Turing rolls and solitons, or can drive the system into a chaotic regime~\cite{cross1993pattern}.

In driven-dissipative optical cavities with Kerr nonlinearity and anomalous dispersion, primary MI leads to coherent optical parametric oscillations (OPO), which are robustly observed across many experimental platforms, including bulk whispering-gallery-mode and integrated photonic microresonators~\cite{savchenkov2004low,herr2012universal}. Theoretical description of optical cavities establishes the spectral dependence of the MI gain rate $\Gamma(\mu)$, where $\mu$ is the mode number, for a given external optical pump power $P_{\text{in}}$ and the detuning of the pump frequency $\omega_p$ from the nearest cavity resonance $\Delta_0$ \cite{haelterman1992dissipative, Kippenberg2004, DelHaye2007, Hansson2013, Matsko2005, Chembo2010, Chembo2013}. For a resonator characterized by an anomalous dispersion coefficient $D_2>0$,  total intracavity losses $\kappa$ and coupling rate $\kappa_{\text{ex}}$ to the pump waveguide, the threshold pump power $P_{\text{th}}$ and the threshold mode $\mu_{\text{th}}$ (that is first involved in the instability development) can be expressed as,
\begin{eqnarray}
    P_{\text{th}} &=& \frac{\hbar\omega_p\kappa^3}{8g_0\kappa_{\text{ex}}} = \frac{n^2_{\text{g}}}{8\omega_p c}\frac{V_{\text{eff}}}{n_{2}}\frac{\kappa^3}{\kappa_{\text{ex}}},\label{eq:threshold}
    \\
    \mu_{\text{th}} &=& \sqrt{\frac{\kappa}{D_2}\biggl ( 1 + \sqrt{ \frac{P_{\text{in}}}{P_{\text{th}}}-1}\biggr )},
\end{eqnarray}
where $g_0 = \frac{\hbar\omega_p^2 c n_2}{n_{\text{g}}^2 V_{\text{eff}}}$ is the Kerr nonlinearity coefficient, $V_{\text{eff}}$ is the effective optical mode volume defined by the geometry of the resonator waveguide, and $n_{\text{g}}$ and \(n_2\) are the group and nonlinear refractive indices.

The OPO produced by primary MI in photonic integrated microresonators enables versatile access to tunable laser wavelengths, promising applications ranging from photon-pair generation to quantum sensing~\cite{brydges2023integrated,black2022optical,guidry2020optical}. Meanwhile, fully developed MI, which involves many unstable modes, underlies the formation of dissipative Kerr solitons (DKS) -- an exceptional source of coherent broadband frequency combs~\cite{Gorodetsky2018dissipative,herr2026frequency}. The mechanism of DKS generation consists of tuning the pump frequency into a strong MI region characterized by chaotic intracavity oscillations, and then exiting this regime into the bistability area, where DKS are formed from MI-produced field fluctuations. Nowadays, using MI to trigger DKS formation is the main approach to producing optical frequency combs in bulk and integrated photonic microresonators, paving the way for miniature devices in frequency metrology, optical telecommunications, and sensing~\cite{Gorodetsky2018dissipative,herr2026frequency}. Thus, understanding MI is one of the key tasks in optical cavity theory.

The classical driven-dissipative Lugiato-Lefever equation (LLE); see \cite{lugiato1987spatial}, successfully captures MI properties in optical cavities, including integrated photonic microresonators~\cite{Gorodetsky2018dissipative,herr2026frequency}. For single optical cavities, linearization of the LLE around a CW background and the study of the corresponding eigenvalue problem enable identification of the MI existence region in the two-dimensional parameter space $(\Delta_0, P_{\text{in}})$, and the analytical description of the MI growth rate in fiber-based, integrated and volume resonators~\cite{haelterman1992dissipative,Chembo2010,Hansson2013,kondratiev2020modulational,anashkina2021optical,Kolesnikova:24}. The emerging interest in the nonlinear regime of coupled microresonator systems along with prominent advances in manufacturing photonic integrated coupled high-Q microresonators raises the fundamental question on theoretical description of MI in such complex multiparametric system \cite{helgason2021dissipative,tikan2021emergent,tikan2022protected,yuan2023soliton,flower2024observation,liu2025near}. While in single ring resonators MI exist only in the case of anomalous dispersion, the coupled systems enable to use of normal dispersion devices in the MI regime thanks to hybridization phenomena with other system components.

\begin{figure}[t!]
    \centering
    \includegraphics[width=1.0\linewidth]{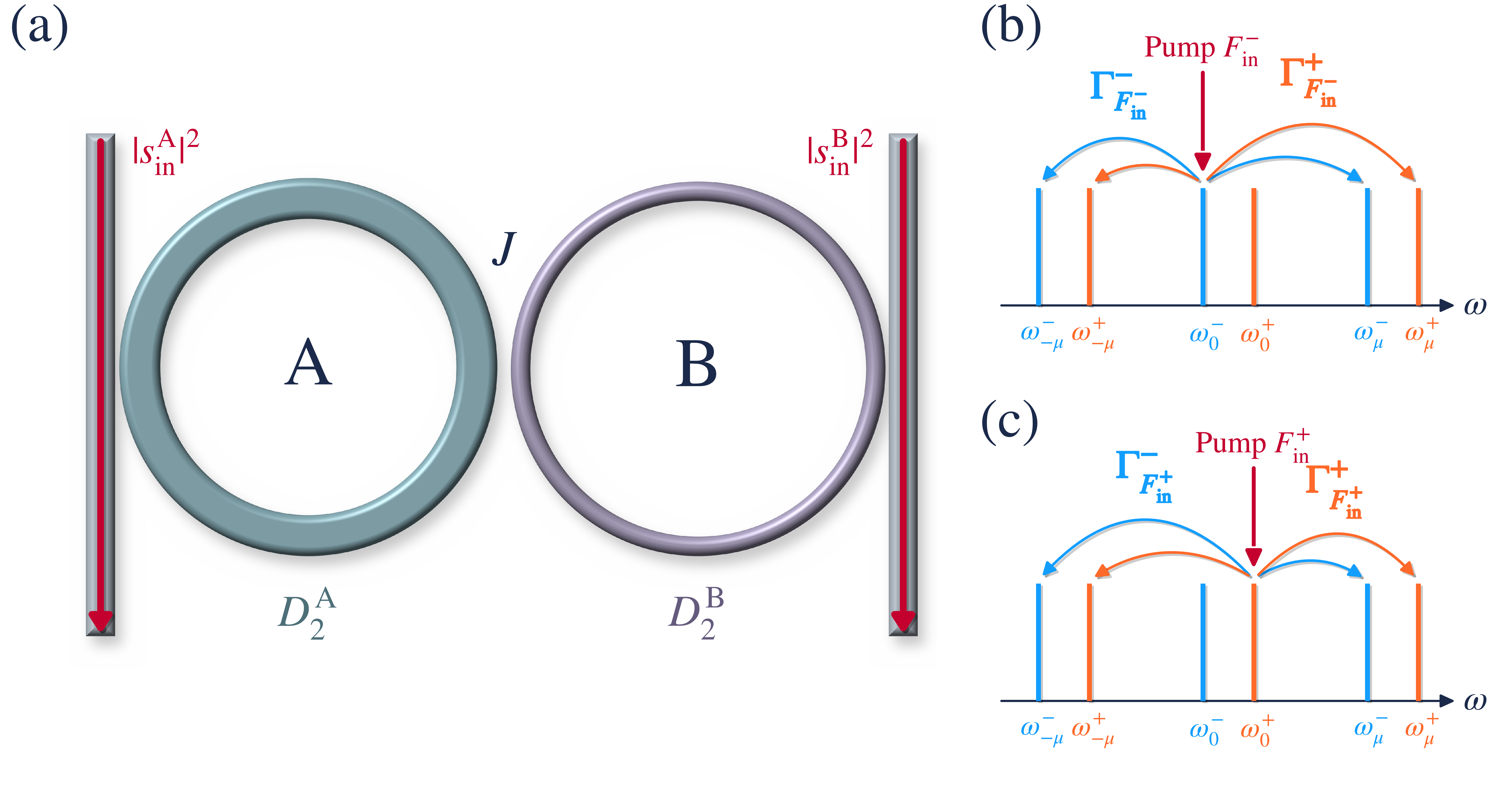}
    \caption{(a) Schematic representation of the coupled microresonators with continuous-wave pump and inter-resonator interaction (\(J>\kappa\)). Schematic of the four-wave-mixing processes for pumping the central mode of the (b) $\omega_0^{-}$ and (c) $\omega_0^{+}$ supermode with pump field $F_{\mathrm{in}}^{\mp}$ (red arrow). Pump photons are converted into sideband pairs at $\omega_{\pm\mu}^{-}$ (blue arrows) with parametric gain $\Gamma^{-}_{F_{\mathrm{in}}}$ and at $\omega_{\pm\mu}^{+}$ (orange arrows) with parametric gain $\Gamma^{+}_{F_{\mathrm{in}}}$, corresponding to intra- and inter-band processes, respectively.}
    \label{fig:1}
\end{figure}

Recent theoretical and experimental studies on various configurations of coupled resonators with anomalous and normal dispersion employ numerical computation of the MI growth rate based on linearized coupled Lugiato-Lefever equations; however, analytical treatment is still lacking \cite{tusnin2023nonlinear,pidgayko2023voltage}. Here we use the Lugiato-Lefever model in the supermode basis and derive a general eigenvalue MI problem for the $4\times 4$ matrix $\mathbf{M}$ corresponding to two coupled resonators with arbitrary second-order dispersion coefficients, arbitrary losses, and coupling strength; see Figure~\ref{fig:1}. We consider the regime of resonance splitting, in which the pump couples only to one of the two supermodes, allowing us to obtain analytical expressions for the stationary continuous-wave field distributed over the coupled system. When the microresonators have similar parameters, the antidiagonal elements of the matrix $\mathbf{M}$ nearly vanish, yielding a fully analytical solution for the four types of the MI gain rate $\Gamma(\mu)$. The four options correspond to the pump and amplification of different supermodes. Finally, we present typical gain rates for coupled anomalous-anomalous, normal-normal, and anomalous-normal microresonators and indicate the cases in which a fully analytical treatment needs to be supplemented by more accurate numerical solutions of the MI eigenvalue problem.

\section{Generalized basis of supermodes}
In the linear regime, the coupled resonators A and B are described using coupled equations for azimuthal mode amplitude \(A_\mu^{\text{i}}\),
\begin{multline}
\label{Eq:coupled-mode_A-B}
    \cfrac{d A_{\mu}^{\text{i}}}{d t}=\left[-\cfrac{\kappa^{\text{i}}}{2}-i\Delta^{\text{i}}-i\cfrac{D_2^{\text{i}}}{2}\mu^2\right]A_\mu^{\text{i}} +iJA_\mu^{\text{j}}\\+\delta_{0\mu}\sqrt{\kappa^{\text{i}}_{\text{c}}}s^{\text{i}}_{\text{in}}, \quad\quad \text{i,\,j}=\{\text{A,B}\},\quad \text{i}\neq \text{j}.
\end{multline}
where \(J\) is the coupling coefficient between microresonators, \(D_2^{\text{i}}\) is a second-order dispersion of microresonator, the intracavity losses are \(\kappa^{\text{i}}\), the coupling coefficient between resonator and input/output waveguide \(\kappa_c^{\text{i}}\) and the pump detunings from microresonators resonant zero-mode frequency: \(\Delta^{\text{A}}=\omega^{\text{A}}_0-\omega_p=\Delta_0+\Delta_{\text{res}}/2\), \(\Delta^{\text{B}}=\omega^{\text{B}}_0-\omega_p=\Delta_0-\Delta_{\text{res}}/2\). The pump vector \(\boldsymbol{s}_{\text{in}}=\left(s_{\text{in}}^{\text{A}},s_{\text{in}}^{\text{B}}\right)\) defines power and phase of the input continuous wave laser field and is related to the total pump power $P_{\text{in}}$ as $|s^{\text{A}}_{\text{in}}|^2 + |s^{\text{B}}_{\text{in}}|^2 = P_{\text{in}}/\hbar\omega_p$.

The resonator's coupling results in the splitting of the azimuthal modes \(\mu\) into a set of supermodes \(\Psi^{\pm}_\mu\) characterized by hybridized dispersion coefficients \(D_{\text{hybr}}^{\pm}(\mu)\) and hybridized spectrally dependent losses \(\kappa_{\text{hybr}}^\pm(\mu)\), following from diagonalization of system (\ref{Eq:coupled-mode_A-B}); resulting in 
generalized coupled mode equations for supermodes:
\begin{equation}
\label{Eq:S1}
    \cfrac{d\Psi^{\pm}_\mu}{d t}=\left[-\cfrac{\kappa_{\text{hybr}}^\pm(\mu)}{2}-i\Delta_{0}-i\cfrac{D_{\text{hybr}}^{\pm}(\mu)}{2}\right]\Psi^{\pm}_\mu+\delta_{0\mu}F^\pm_{\text{in}}.
\end{equation}
Here, the hybridized dispersion and spectrally dependent losses were found from the solution of the eigenvalue problem of the initial matrix of coupled equations:
\begin{eqnarray}
\label{Eq:Spl:Dhybr}
D_{\text{hybr}}^\pm(\mu) &=& \cfrac{D_2^{\text{A}}+D_2^{\text{B}}}{2}\mu^2\mp\cfrac{\text{sgn}(y)}{\sqrt{2}}\left(\sqrt{x^2+y^2}-x\right)^{1/2}
\\
\kappa_{\text{hybr}}^\pm(\mu) &=& \cfrac{\kappa^{\text{A}}+\kappa^{\text{B}}}{2}\mp\cfrac{\text{sgn}(y)}{\sqrt{2}}\left(\sqrt{x^2+y^2}+x\right)^{1/2},
\end{eqnarray}
where the coefficients $x$ and $y$ are,
\begin{eqnarray}
    x &=& \cfrac{\left(\kappa^{\text{A}}-\kappa^{\text{B}}\right)^2}{4}-\left(\Delta_{\text{res}}+\cfrac{D_2^{\text{A}}-D_2^{\text{B}}}{2}\mu^2\right)^2-4J^2
    \\
    y &=& \left(\kappa^{\text{A}}-\kappa^{\text{B}}\right)\left(\Delta_{\text{res}}+\cfrac{D_2^{\text{A}}-D_2^{\text{B}}}{2}\mu^2\right).
\end{eqnarray}
The expression of supermodes (eigenvectors) and pump vector in terms of the basis of resonator modes also follows from the solution of the eigenvalue problem and can be written as,
\begin{eqnarray}
\label{Eq:Spl:direct-tranformation:A-to-Psi}
\Psi^{\pm}_\mu &=& \cfrac{A_\mu^{\text{A}}+C^{\pm}_{\mu} A_\mu^{\text{B}}}{\sqrt{1+|C^{\pm}_{\mu}|^2}}\,,
\\
\label{Eq:Spl:direct-tranformation:s-to-F}
F_{\text{in}}^\pm &=& \cfrac{\sqrt{\kappa_c^{A}}s^{A}_{\text{in}}+C^\pm_0 \sqrt{\kappa_c^{B}}s^{B}_{\text{in}}}{\sqrt{1+|C^\pm_0|^2}}\,,
\\
A^{\text{A}}_\mu &=& \cfrac{C^-_{\mu}\sqrt{1+|C^+_{\mu}|^2}\Psi^{+}_\mu-C^+_{\mu} \sqrt{1+|C^-_{\mu}|^2}\Psi^{-}_\mu}{C^-_{\mu}-C^+_{\mu}},
\label{Eq:Spl:inverse-tranformation:Psi-to-A}
\\
A^{\text{B}}_\mu &=& \cfrac{-\sqrt{1+|C^+_{\mu}|^2}\Psi^{+}_\mu+ \sqrt{1+|C^-_{\mu}|^2}\Psi^{-}_\mu}{C^-_{\mu}-C^+_{\mu}},
\label{Eq:Spl:inverse-tranformation:Psi-to-B}
\end{eqnarray}
where
\begin{multline}
\label{Eq:Spl:Cmu}
    C^{\pm}_{\mu} =  J^{-1}\left(i\cfrac{\kappa^{\text{B}}-\kappa^{\text{A}}}{4}+\cfrac{\Delta_{\text{res}}}{2}+\cfrac{D_2^{\text{A}}-D_2^{\text{B}}}{4}\mu^2\pm\right.\\\left.\cfrac{1}{2\sqrt{2}}\left(-i\left(\sqrt{x^2+y^2}+x\right)^{1/2}+\left(\sqrt{x^2+y^2}-x\right)^{1/2}\right)\right).
\end{multline}

\section{Nonlinear theory}
To study the nonlinear dynamics of our coupled microresonator system, we use the Lugiato-Lefever model written for the intracavity field distribution $A^{\text{i}}(\theta,t)=\sum_{\mu\in\mathbb{Z}} A^{\text{i}}_{\mu}(t)e^{-i\mu\theta}$ as follows:
\begin{multline}
\label{Eq:coupled-LLE}
    \cfrac{\partial A^{\text{i}}}{\partial t}=\left[-\cfrac{\kappa^{\text{i}}}{2}-i\Delta^{\text{i}}+i\cfrac{D_2^{\text{i}}}{2}\cfrac{\partial^2}{\partial\theta^2}+ig^{\text{i}}\left|A^{\text{i}}\right|^2\right]A^{\text{i}} \\+iJA^{\text{j}}+\sqrt{\kappa^{\text{i}}_c}s_{\text{in}}^{\text{i}}, \quad\quad
    \text{i},\text{j} =\{\text{A,B}\},\quad \text{i}\neq\text{j}.
\end{multline}
 where \(g^{\text{i}}\) are the nonlinear coefficients.
 
 Before proceeding to solve the modulation instability problem, it is necessary to address the pump mode field. It is important to note that we are considering a problem involving single-frequency pumping, in which the pump can be applied to both resonators. Since the microresonator coupling splits modes into a set of supermodes when \(J>\kappa\), we can selectively pump one of them, see Figure~\ref{fig:1}. Thus, the ratio of pump amplitudes \(s^{\text{i}}_{\text{in}}\) into the different resonators determines only the efficiency with which the supermode is pumped, while the choice of which supermode is pumped is determined by the pump frequency detuning \(\Delta^{\text{i}}\).
 
\subsection{Pump mode}
To find a solution for the pump mode, we consider an approximation in which only one mode exists in the system with \(\mu=0\). Substituting the ansatz $A^{\text{i}}(\theta,t)= A^{\text{i}}_{0}(t)$ into equation \eqref{Eq:coupled-LLE}, we obtain:
\begin{multline}
\label{Eq:pump-mode_A-B}
    \cfrac{d A_{0}^{\text{i}}}{d t}=\left[-\cfrac{\kappa^{\text{i}}}{2}-i\Delta^{\text{i}}\right]A_0^{\text{i}} +iJA_0^{\text{j}}+ig^{\text{i}}|A^\text{i}_0|^2A_{0}^{\text{i}}\\+\sqrt{\kappa^{\text{i}}_{\text{c}}}s^{\text{i}}_{\text{in}}, \quad\quad \text{i},\text{j}=\{\text{A,B}\},\quad \text{i}\neq\text{j}.
\end{multline}

The system of nonlinear equations for pumping mode \(\mu\) in the supermode basis \(\Psi^{\pm}_0=\cfrac{A_0^{\text{A}}+C^\pm_0 A_0^{\text{B}}}{\sqrt{1+|C^\pm_0|^2}}\) reads as:
 \begin{widetext}
\begin{multline}
    \cfrac{d \Psi^{\pm}_0}{d t}=\left[-\cfrac{\kappa^\pm_{\text{hybr}}(0)}{2}-i\Delta_{0}-i\cfrac{D_{\text{hybr}}^{\pm}(0)}{2}\right]\Psi^{\pm}_0+F^\pm_{\text{in}}+\\+\cfrac{ig^{\text{A}}}{\sqrt{1+|C^\pm_0|^2}}\left|\cfrac{C^-_0\sqrt{1+|C^+_0|^2}\Psi^{+}_0-C^+_0\sqrt{1+|C^-_0|^2}\Psi^{-}_0}{C^-_0-C^+_0}\right|^2\left(\cfrac{C^-_0\sqrt{1+|C^+_0|^2}\Psi^{+}_0-C^+_0\sqrt{1+|C^-_0|^2}\Psi^{-}_0}{C^-_0-C^+_0}\right)+\\+\cfrac{ig^{\text{B}}C^\pm_0}{\sqrt{1+|C^\pm_0|^2}}\left|\cfrac{-\sqrt{1+|C^+_0|^2}\Psi^{+}_0+\sqrt{1+|C^-_0|^2}\Psi^{-}_0}{C^-_0-C^+_0}\right|^2\left(\cfrac{-\sqrt{1+|C^+_0|^2}\Psi^{+}_0+\sqrt{1+|C^-_0|^2}\Psi^{-}_0}{C^-_0-C^+_0}\right).
\end{multline}
 \end{widetext}
 If the amplitudes and phases of the pump field are chosen optimally, we can efficiently excite only one of the supermodes~--~the one whose frequency is close to that of the radiation source~--~in which case \(F^\pm_{\text{in}}\neq0\), \(F^\mp_{\text{in}}=0\) (\(\Psi_0^\mp\equiv0\)), then the equations are separated, and we obtain independent equations for the two cases of pumping into the \(\Psi^{+}_0\) or \(\Psi^{-}_0\) :
 \begin{widetext}
\begin{multline}
\label{eq:psi^pm_0}
    \cfrac{d \Psi^{\pm}_0}{d t}=\left[-\cfrac{\kappa^\pm_{\text{hybr}}(0)}{2}-i\Delta_{0}-i\cfrac{D_{\text{hybr}}^{\pm}(0)}{2}\right]\Psi^{\pm}_0+F^\pm_{\text{in}}+i\cfrac{\pm\left(g^{\text{A}}|C^\mp_0|^2C^\mp_0-g^{\text{B}}C^\pm_0\right)}{|C^-_0-C^+_0|^2\left(C^-_0-C^+_0\right)}(1+|C^\pm_0|^2)\left|\Psi^{\pm}_0\right|^2\Psi^{\pm}_0.
\end{multline}
 \end{widetext}
 Remarkably, even if the amplitudes are not chosen optimally, the power coupled into another supermode will be negligible due to the large frequency detuning of the radiation source. The equations for the pump mode in the supermode basis coincide in form with the equation for a single microresonator, up to the recalculation of the effective nonlinearity coefficient \(G_0^\pm=\cfrac{\pm\left(g^{\text{A}}|C^\mp_0|^2C^\mp_0-g^{\text{B}}C^\pm_0\right)}{|C^-_0-C^+_0|^2\left(C^-_0-C^+_0\right)}(1+|C^\pm_0|^2)\), which is necessary due to the asymmetry of the field distribution in the two resonators. So, the stationary solution of (\ref{eq:psi^pm_0}) is:
\begin{multline}
    \left(\cfrac{F^\pm_{\text{in}}}{|\Psi^\pm_0|}\right)^2-\left(\cfrac{\kappa^\pm_{\text{hybr}}(0)}{2}+\text{Im}(G_0^\pm)|\Psi_0^\pm|^2\right)^2=\\=\left(\text{Re}\left(G_0^\pm\right)|\Psi_0^\pm|^2-\Delta_0-D^\pm_{\text{hybr}}(0)/2\right)^2,\\
    \Psi^\pm_0=\cfrac{F^\pm_{\text{in}}}{\cfrac{\kappa^\pm_{\text{hybr}}(0)}{2}+i\Delta_{0}+i\cfrac{D_{\text{hybr}}^{\pm}(0)}{2}-iG_0^\pm|\Psi^\pm_0|^2}.
\end{multline}

\begin{figure}[htpb!]
    \centering
\includegraphics[width=\linewidth]{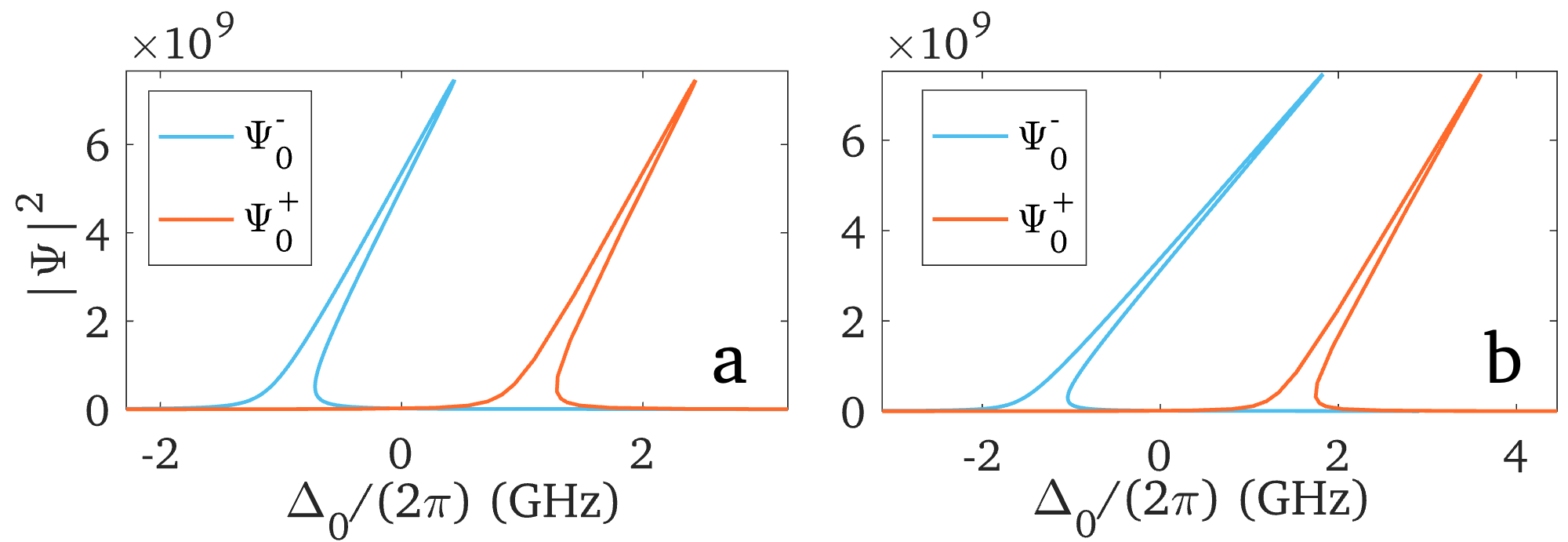}
    \caption{Resonance Kerr tilt of the CW solution for pumped \(\Psi^-_0\)/\(\Psi^+_0\) supermodes a) in the same microresonators and b) in microresonators with different sign of dispersion in the case of strong couppling \(J>\kappa\).}
    \label{fig:resonances}
\end{figure}

Note that in the case of identical losses \(\kappa^{\text{A}}=\kappa^{\text{B}}\), \(\text{Im}(G_0^\pm)\equiv0\), and solutions for pumping supermodes look like solutions for a single microresonator (Fig.\ref{fig:resonances}). In this regard, the threshold power for coupled microresonators is defined as in equation~\eqref{eq:threshold}, subject to a substitution \(g_0\rightarrow G_0^\pm\):
\begin{eqnarray}
    P_{\text{th}}^\pm &=& \frac{\hbar\omega_p\kappa^3}{8G_0^\pm\kappa_{\text{ex}}} .\label{eq:general_threshold}
\end{eqnarray}

For identical coupled microresonators, the CW solution for pumping into different supermodes is the same (Fig.\ref{fig:resonances}a); however, for resonators with different volumes, the effective nonlinearity coefficients \(G_0^+\neq G_0^-\), which leads to different slopes for the nonlinear resonances (Fig.\ref{fig:resonances}b).

\subsection{Modulation instability analysis}
Once we have obtained the solution to the pump-mode equation, we can proceed to address the modulation instability of the sidebands relative to the pump mode. For this purpose, we substitute the field ansatz $A^{\text{i}}(\theta,t)=A^{\text{i}}_{0}(t)+A^{\text{i}}_{\mu}(t)e^{-i\mu\theta}+A^{\text{i}}_{-\mu}(t)e^{i\mu\theta}$ in equation \eqref{Eq:coupled-LLE} and linearize it under the assumption of small noise-like amplitudes of the side modes:
 \begin{widetext}
\begin{equation}
    \begin{cases}

    \cfrac{d A^\text{A}_\mu}{d t}=\left[-\cfrac{\kappa^\text{A}}{2}-i\Delta^{\text{A}}-iD_2^{\text{A}}\mu^2/2+i2g^{\text{A}}\left|A^\text{A}_0\right|^2\right]A^\text{A}_\mu +ig^{\text{A}}(A^\text{A}_0)^2(A^{\text{A}}_{-\mu})^*+iJA^\text{B}_\mu ,\\

 \cfrac{d (A_{-\mu}^\text{A})^*}{d t}=\left[-\cfrac{\kappa^\text{A}}{2}+i\Delta^{\text{A}}+iD_2^{\text{A}}\mu^2/2-i2g^{\text{A}}\left|A^\text{A}_0\right|^2\right](A^\text{A}_{-\mu})^* -ig^{\text{A}}(A^\text{A}_0)^{*2}A^\text{A}_{\mu}-iJ(A^\text{B}_{-\mu})^* ,\\
 
    \cfrac{d A^\text{B}_\mu}{d t}=\left[-\cfrac{\kappa^\text{B}}{2}-i\Delta^{\text{B}}-iD_2^{\text{B}}\mu^2/2+i2g^{\text{B}}\left|A^\text{B}_0\right|^2\right]A^\text{B}_\mu +ig^{\text{B}}(A^\text{B}_0)^2(A^\text{B}_{-\mu})^{*}+iJA^\text{A}_\mu ,\\

 \cfrac{d (A^\text{B}_{-\mu})^*}{d t}=\left[-\cfrac{\kappa^\text{B}}{2}+i\Delta^{\text{B}}+iD_2^{\text{B}}\mu^2/2-i2g^{\text{B}}\left|A^\text{B}_0\right|^2\right](A^\text{B}_{-\mu})^* -ig^{\text{B}}(A^\text{B}_0)^{*2}A^\text{B}_{\mu}-iJ(A^\text{A}_{-\mu})^* .\\\label{Eq_syst_microresonators}
   \end{cases}
\end{equation}
\end{widetext}
We will consider the case when pumping is carried out only in one supermode. Then for the pumping into supermode  \(\Psi^-_0\) \(A^\text{A}_0 = \cfrac{-C^+_0\sqrt{1+|C^-_0|^2}}{C^-_0-C^+_0}\Psi^-_0\), \(A^\text{B}_0=\cfrac{\sqrt{1+|C^-_0|^2}}{C^-_0-C^+_0}\Psi^-_0\), and for the supermode \(\Psi^+_0\) \(A^\text{A}_0 = \cfrac{C^-_0\sqrt{1+|C^+_0|^2}}{C^-_0-C^+_0}\Psi^+_0\), \(A^\text{B}_0=\cfrac{-\sqrt{1+|C^+_0|^2}}{C^-_0-C^+_0}\Psi^+_0\). In the supermode basis the system  \eqref{Eq_syst_microresonators} can be represented as:

\begin{equation}
    \cfrac{d}{dt}\begin{pmatrix}
        \Psi^-_\mu\\
        \Psi^{-*}_{-\mu}\\
        \Psi^{+}_{\mu}\\
        \Psi^{+*}_{-\mu}\\
    \end{pmatrix}
    =\underbrace{\begin{pmatrix}
        \text{A}^-& \text{B}^- & \text{C}^- & \text{D}^-\\
        \text{B}^{-*}& \text{A}^{-*}& \text{D}^{-*}& \text{C}^{-*}\\
         \text{C}^+ & \text{D}^+&\text{A}^+& \text{B}^+ \\
       \text{D}^{+*}& \text{C}^{+*}&\text{B}^{+*}& \text{A}^{+*}\\
    \end{pmatrix}}_{\mathbf{M}}
    \begin{pmatrix}
        \Psi^-_\mu\\
        \Psi^{-*}_{-\mu}\\
        \Psi^{+}_{\mu}\\
        \Psi^{+*}_{-\mu}\\
    \end{pmatrix}.\label{General_MI}
\end{equation}
Here the elements of matrix $\mathbf{M}$ are:
 \begin{widetext}
 \begin{multline}
     \text{A}^\pm = -\cfrac{\kappa^\pm_{\text{hybr}}(\mu)}{2}-i\Delta_{0}-i\cfrac{D_{\text{hybr}}^{\pm}(\mu)}{2}+\cfrac{2ig^{\text{B}}|A^\text{B}_0|^2C^\pm_\mu-2ig^{\text{A}}|A^\text{A}_0|^2C^\mp_\mu}{C^\pm_\mu-C^\mp_\mu}; \text{B}^\pm = \cfrac{ig^{\text{A}}(A^\text{A}_0)^2C^{\mp *}_\mu-ig^{\text{B}}(A^\text{B}_0)^2C^\pm_\mu}{C^{\mp *}_\mu-C^{\pm *}_\mu};\\
     \text{C}^\pm= \cfrac{i2C^\pm_\mu\sqrt{1+|C^\mp_\mu|^2}}{\sqrt{1+|C^\pm_\mu|^2}(C^\mp_\mu-C^\pm_\mu)}\left(g^{\text{B}}|A^\text{B}_0|^2-g^{\text{A}}|A^\text{A}_0|^2\right);\text{D}^\pm=\cfrac{i\sqrt{1+|C^\mp_\mu|^2}}{\sqrt{1+|C^\pm_\mu|^2}(C^{\mp *}_\mu-C^{\pm *}_\mu)}\left(C^\pm_\mu g^{\text{B}}(A^\text{B}_0)^2-C^{\pm *}_\mu g^{\text{A}}(A^\text{A}_0)^2\right).
 \end{multline}
 \end{widetext}
 

To address the modulation instability problem we assume \(\Psi^\pm_\mu \sim \exp(\lambda t)\) and 
substitute it into Eq.~(\ref{General_MI}), which results in the eigenvalue problem for matrix $\mathbf{M}$.
Then, the parametric gain coefficients of the sidemodes can be found via real part of the eigenvalues $\lambda$ as
\begin{equation}
    \Gamma^\pm_{F^\pm_{\text{in}}}(\mu,\Delta_0,|A^i_0|^2)=\text{Re}\{\lambda^\pm_{F^\pm_{\text{in}}}(\mu)\}.
\end{equation}

In general, the eigenvalue problem for matrix $\mathbf{M}$ can be solved only numerically. However, in the case of microresonators with similar geometry \(\cfrac{g^{\text{A}}|A_0^{\text{A}}|^2}{g^{\text{B}}|A_0^{\text{B}}|^2}\rightarrow1\),  equations (\ref{General_MI}) are separated by supermodes, i.e., the mutual contribution between the supermodes can be neglected (\(\text{C}^\pm=0\), \(\text{D}^\pm=0\)). For different microresonators, the term \(\left|g^{\text{B}}|A^\text{B}_0|^2C^\mp_\mu-g^{\text{A}}|A^\text{A}_0|^2C^\pm_\mu\right|\gg \left|g^{\text{B}}|A^\text{B}_0|^2-g^{\text{A}}|A^\text{A}_0|^2\right|\) is due to the opposite signs of the real parts \(C^-_\mu\) and \(C^+_\mu\). Since the contribution of the counterpart supermode is small, it is possible, in the first approximation, to solve the system of equations for the supermodes independently. In this case, the analysis is similar to that of modulation instability in a single resonator with hybridized dispersion and nonlinearity. Thus, in the first approximation, matrix $\mathbf{M}$ can be treated as block-diagonal leading to analytical expression for its eigenvalues.
In the case of equal losses \(\kappa^{\text{A}}=\kappa^{\text{B}}\equiv\kappa\), the decrements have the form, which is similar to that of a single microresonator. For pumping into supermode \(\Psi^-_0\), positive increment for supermodes \(\Psi^\pm_\mu\) is defined by
\begin{multline} 
    \lambda^{\pm}_{F^-_{\text{in}}}(\mu)=-\cfrac{\kappa}{2}+\\\sqrt{G_{F^-_{\text{in}}}^{\pm 2}(\mu)|\Psi^-_0|^4-\left(\Delta_0+\cfrac{D_{\text{hybr}}^\pm(\mu)}{2}-2G^{\pm}_{F^-_{\text{in}}}(\mu)|\Psi^-_0|^2\right)^2}; \\
    G_{F^-_{\text{in}}}^\pm(\mu) = (1+|C^-_0|^2)\cfrac{g^{\text{A}}C_\mu^\mp-g^{\text{B}}(C_0^+)^2C_\mu^\pm}{(C_\mu^\mp-C_\mu^\pm)(C_0^--C_0^+)^2}.\label{eq:gain}
\end{multline}

For pumping into supermode \(\Psi^+_0\), an analogous positive increment for supermodes \(\Psi^\pm_\mu\) reads as
\begin{multline} 
    \lambda^{\pm}_{F^+_{\text{in}}}(\mu)=-\cfrac{\kappa}{2}+\\\sqrt{G_{F^+_{\text{in}}}^{\pm 2}(\mu)|\Psi^+_0|^4-\left(\Delta_0+\cfrac{D_{\text{hybr}}^\pm(\mu)}{2}-2G^{\pm}_{F^+_{\text{in}}}(\mu)|\Psi^+_0|^2\right)^2}; \\
    G_{F^+_{\text{in}}}^\pm(\mu) =(1+|C_0^+|^2) \cfrac{g^{\text{A}}(C_0^{-})^2C_\mu^\mp-g^{\text{B}}C_\mu^\pm}{(C^\mp_\mu-C^\pm_\mu)(C_0^--C_0^+)^2.}\label{gain2}
\end{multline}


The increment for the case of different losses is considered in the Appendix. In the next section, we will examine several examples to demonstrate the accuracy of the analytical approximation for the case of equal losses \(\kappa^{\text{A}}=\kappa^{\text{B}}\equiv\kappa\).
\section{Specific examples}

\subsection{Equal microresonators}
In the case of identical microresonators \(g^{\text{A}}=g^{\text{B}}\equiv g\), \(\Delta^{\text{A}}=\Delta^{\text{B}}\equiv \Delta_0\), \(D_2^{\text{A}}=D_2^{\text{B}}\equiv D_2\) the system of equations \eqref{General_MI} is divided into symmetric (S) \(\Psi^+_\mu=\cfrac{A^\text{A}_\mu+A^\text{B}_\mu}{\sqrt{2}}\) and antisymmetric (AS) \(\Psi^-_\mu=\cfrac{A^\text{A}_\mu-A^\text{B}_\mu}{\sqrt{2}}\)  supermodes exactly (\(\text{C}^\pm=0\), \(\text{D}^\pm=0\)). Each resonance splits into two with \(2J\) offset (\(D_\text{hybr}^\pm/2=D_2\mu^2/2\mp J\)). So we have an exact analytical expression for the modulation instability positive increment:
\begin{widetext} 
\begin{eqnarray}
    \lambda^{\pm}_{\text{AS}}(\mu)&=&-\cfrac{\kappa}{2}+\sqrt{g^2|\Psi^-_0|^4/4-\left(\Delta_0+D_2\mu^2/2\mp J-g|\Psi^-_0|^2\right)^2},\,\,\,\,\text{pumping in AS mode } \Psi_0^-\\
    \lambda^{\pm}_{\text{S}}(\mu)&=&-\cfrac{\kappa}{2}+\sqrt{g^2|\Psi^+_0|^4/4-\left(\Delta_0+D_2\mu^2/2\mp J-g|\Psi^+_0|^2\right)^2},\,\,\,\, \text{pumping in S mode } \Psi_0^+.
\end{eqnarray}
\end{widetext}

\begin{figure}[!t]
    \includegraphics[width=\linewidth]{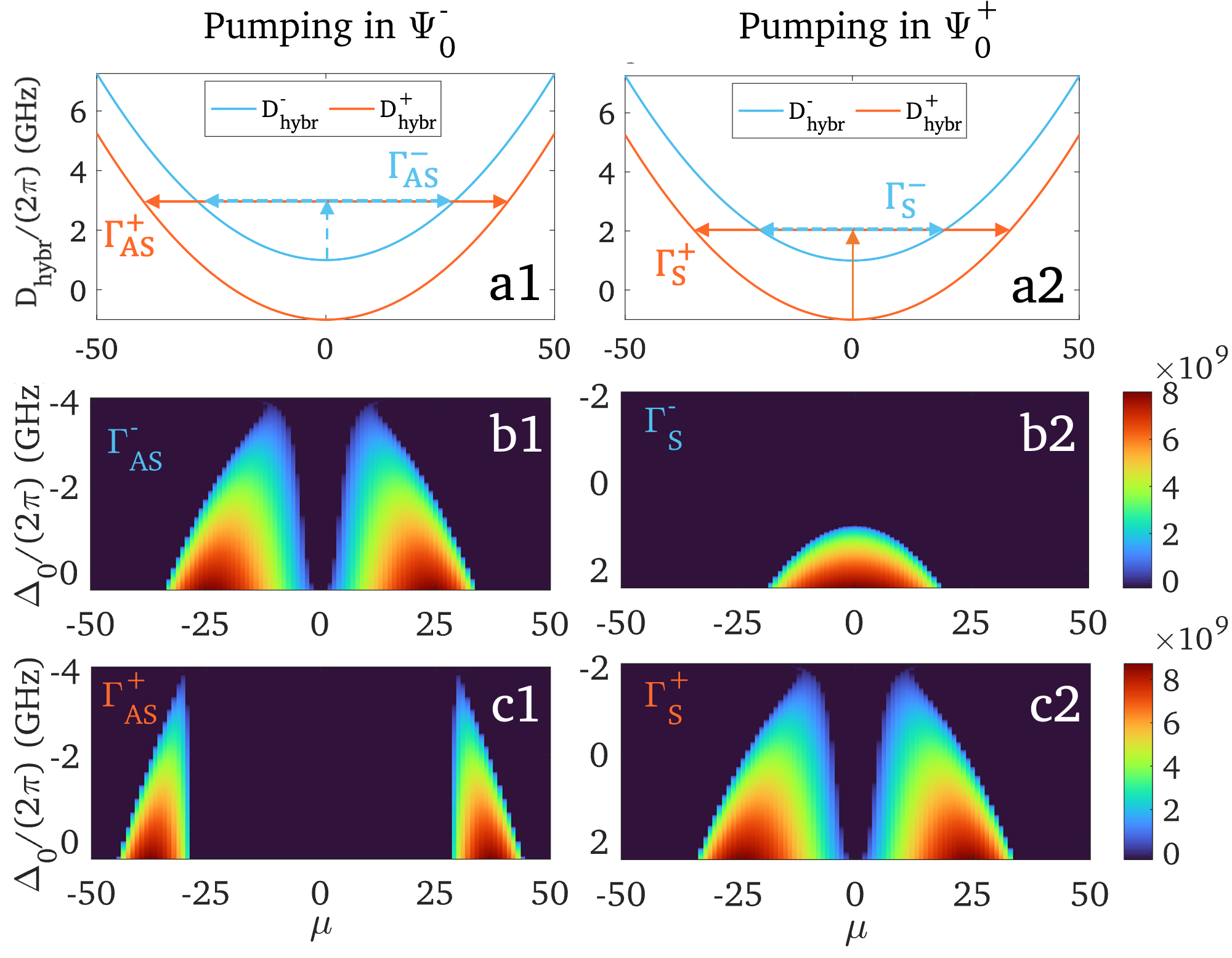}
    \caption{(a1, a2) Integrated dispersion of the hybridized supermode families for the same anomalous dispersion of microresonators \(D_2/(2\pi)=5\) MHz; arrows indicate the four-wave mixing processes for pumping in $\Psi_0^{-}$ (a1) and $\Psi_0^{+}$ (a2). (b1, c1) MI gain $\Gamma^{-}_{\text{AS}}$ and $\Gamma^{+}_{\text{AS}}$ versus $\mu$ and pump detuning $\Delta_0/2\pi$ for pumping in $\Psi_0^{-}$. (b2, c2) MI gain $\Gamma^{-}_{\text{S}}$ and $\Gamma^{+}_{\text{S}}$ for pumping in $\Psi_0^{+}$ (\(\kappa = 100\) MHz, \(P_{\text{in}} = 0.3\) W).}\label{fig:anom-anom}
\end{figure}

From the theory of modulation instability for a single microresonator, we know that positive modal dispersion is the critical quantity for the occurrence of parametric generation. Thus, in the case of two identical anomalous rings, there is no dispersion hybridization, and the levels are simply split into two (see Fig.\ref{fig:anom-anom}a).
It can be seen that when pumping into symmetric/antisymmetric modes, the gain lobes of the side symmetric/antisymmetric modes \(\mu\) look the same as in a single microresonator (see Fig. \ref{fig:anom-anom} b1 and c2). In addition, it can be seen from Figure \ref{fig:anom-anom}a1 that the detuning of the resonant frequency of the antisymmetric pump supermode \(\mu=0\) is compensated by the dispersion of the symmetric supermodes with number \(\mu
\sim\pm 35\) (in Figure \ref{fig:anom-anom}a1, the resonances lie on the same horizontal line). Therefore, by pumping an antisymmetric supermode, it is possible to excite symmetric side modes as well (see Fig. \ref{fig:anom-anom}c1). On the other hand, at large detunings of the symmetric supermode, we enter the region of modulation instability of antisymmetric side modes with \(\mu\) near zero (see Fig. \ref{fig:anom-anom}b2).

\begin{figure}[htp!]
    \centering
    \includegraphics[width=\linewidth]{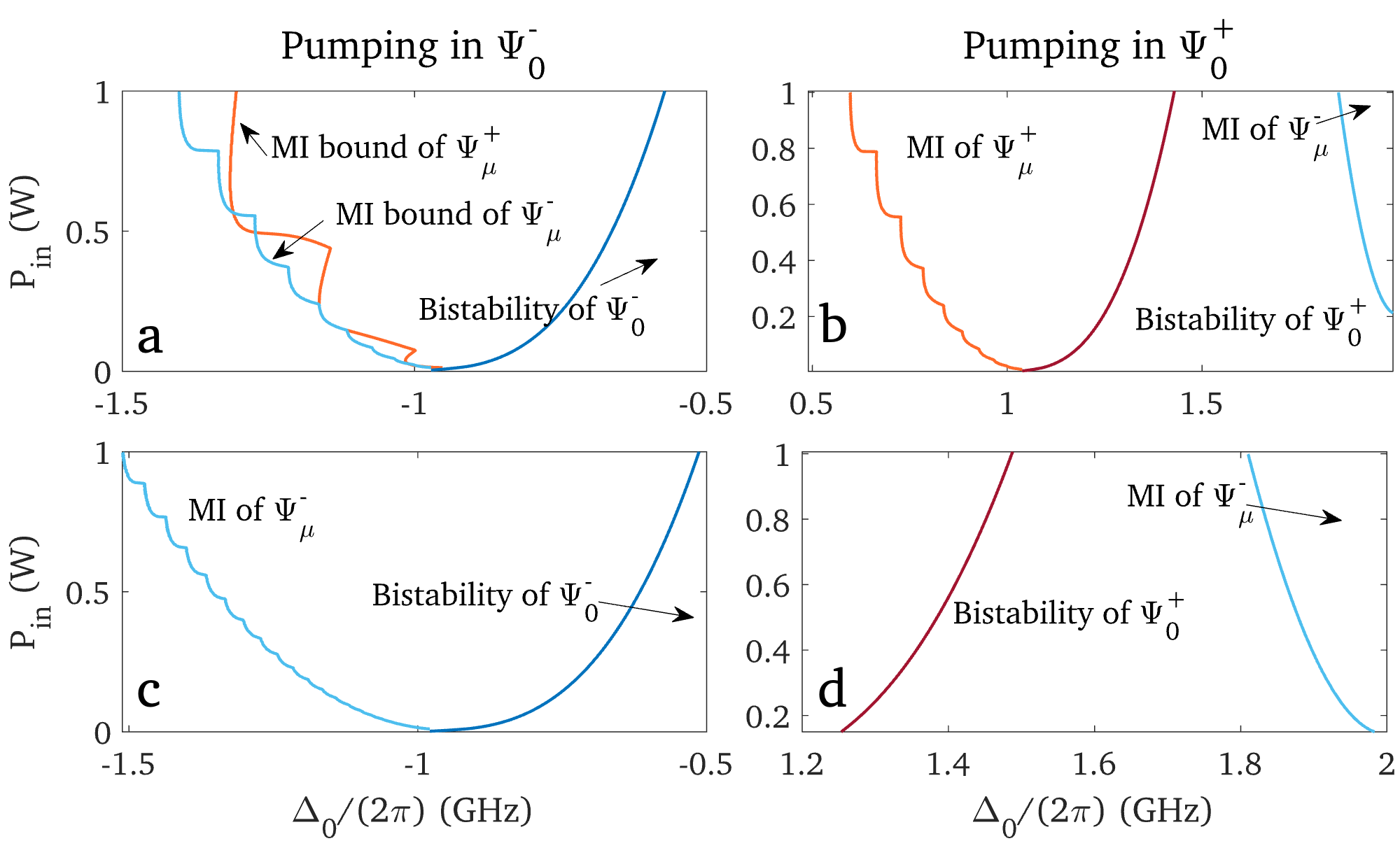}
    \caption{The boundary of the modulation instability region in the two-dimensional parameter space $(\Delta_0, P_{\text{in}})$ a, b) in the case of identical microresonators with anomalous dispersion \(D_2/(2\pi)=5\) MHz and c, d) in the case of microresonators with opposite sign of dispersion \(D_2^{\text{A}}/(2\pi)=1\) MHz and \(D_2^{\text{B}}/(2\pi)=-1\) MHz.}
    \label{fig:mi_region}
\end{figure}

The analytical expression for the increments \eqref{eq:gain}, \eqref{gain2} also makes it possible to obtain the regions of modulation instability on the phase diagram (\(\Delta_0, P_{\text{in}}\)) (see Fig. \ref{fig:mi_region}). For pumping into \(\Psi_0^-\) we observe that the regions of modulation instability for \(\Psi_\mu^-\) and \(\Psi_\mu^+\) supermodes are essentially identical. Conversely, for the pumping of \(\Psi_0^+\), we observe a pattern of the modulation instability region of \(\Psi_\mu^+\) that is typical for a single microresonator; furthermore, the modulation instability region for \(\Psi_\mu^-\) lies far from the boundary of the mode \(\Psi_0^+\) bistability region (see Fig. \ref{fig:mi_region}b).

\begin{figure}[htp!]
    \centering
    \includegraphics[width=\linewidth]{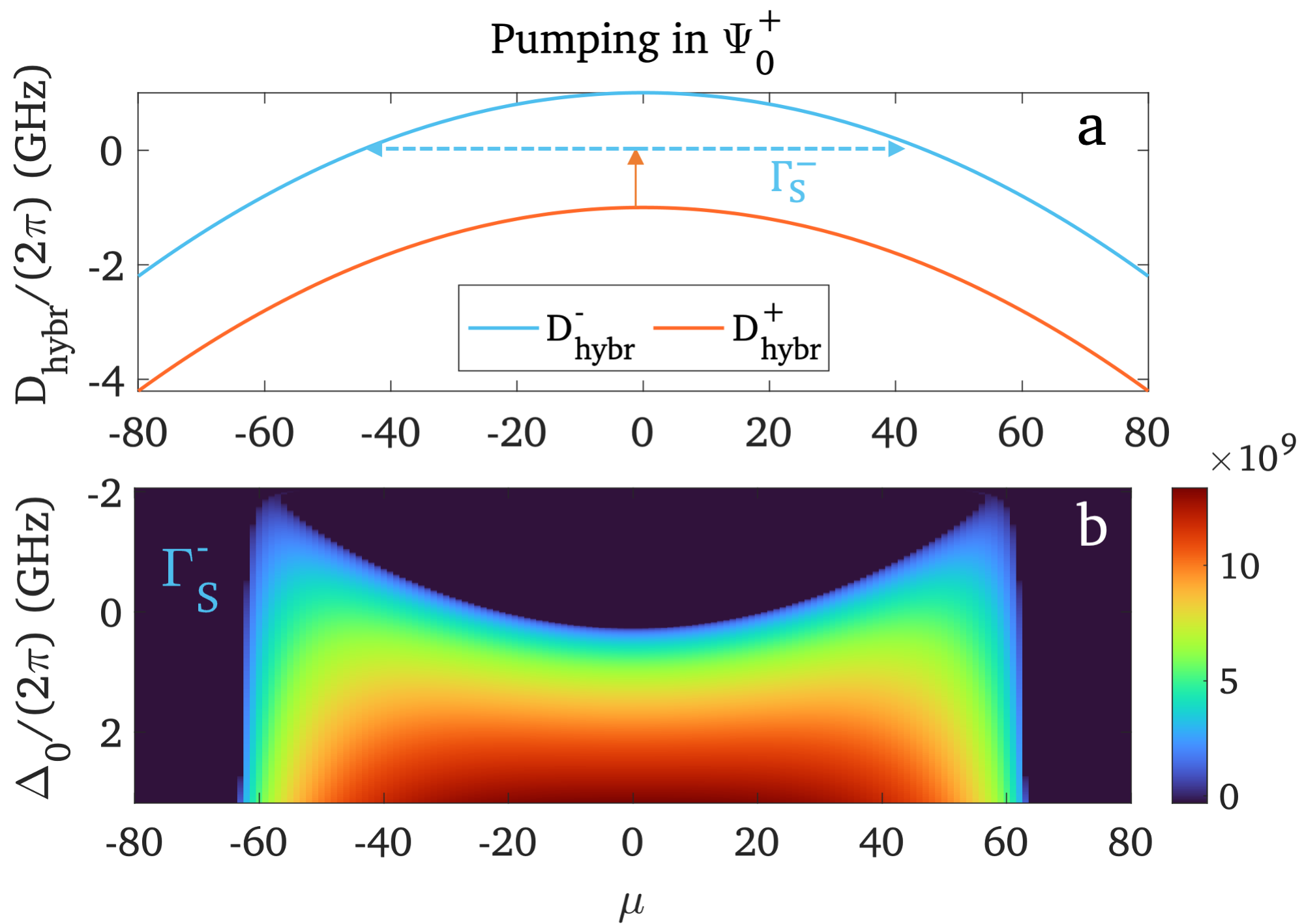}
    \caption{(a) Integrated dispersion of the hybridized supermode families for the same normal dispersion of microresonators \(D_2/(2\pi)=-1\) MHz; arrows indicate the four-wave mixing processes for pumping in $\Psi_0^{+}$. (b) MI gain $\Gamma^{-}_{\text{S}}$ versus $\mu$ and pump detuning $\Delta_0/2\pi$ for pumping in $\Psi_0^{+}$ (\(\kappa = 100\) MHz, \(P_{\text{in}} = 0.3\) W).}
    \label{fig:norm-norm}
\end{figure}

In the case of coupled identical microresonators with normal dispersion, the splitting of the levels leads to the fact that the detuning of the symmetric supermode is compensated by the dispersion of the antisymmetric side modes (see Fig.\ref{fig:norm-norm}a, the symmetric modes with \(\mu=0\) lie on the same horizontal line as the antisymmetric modes with \(\mu\sim\pm 70\)). Thus, in such a system, modulation instability of antisymmetric side modes is possible, arising from pumping into a symmetric supermode. Since the dispersion is generally normal, no other four-wave mixing processes occur.

\subsection{Anomalous-normal dispersion}
For different microresonators, the system of equations \eqref{eq:gain} becomes inaccurate. In this section, we test how well the analytical expressions predict the exact solution for the system \eqref{General_MI}. Recall that the degree of applicability of an analytical solution is determined by the value of \(\cfrac{g^{\text{A}}|A^\text{A}_0|^2}{g^{\text{B}}|A^\text{B}_0|^2}\). As follows from Equations \eqref{Eq:Spl:inverse-tranformation:Psi-to-A}, \eqref{Eq:Spl:inverse-tranformation:Psi-to-B} this relationship is proportional to \(|C_0^\pm|\) in the case where only one supermode \(\Psi_0^\mp\) is pumped. Thus, for the analysis to remain valid, it is sufficient that the coefficients \(g^{\text{A}}/g^{\text{B}}|C_0^\pm|^2\sim1\).
Thus, for various microresonators, given identical losses, \(C_0^\pm=\left(\Delta_{\text{res}}/2\pm\sqrt{\Delta_{\text{res}}^2/4+J^2}\right)/J\), and for absolute detuning of the resonances \(\Delta_{\text{res}}\ll J\) and a small difference in the volumes of the resonators (\(g^{\text{A}}/g^{\text{B}}\sim 1\)), we get \(g^{\text{A}}/g^{\text{B}}|C_0^\pm|\sim1\); consequently, the analytical approach yields precise results.

\begin{figure}[t!]
    \centering
    \includegraphics[width=\linewidth]{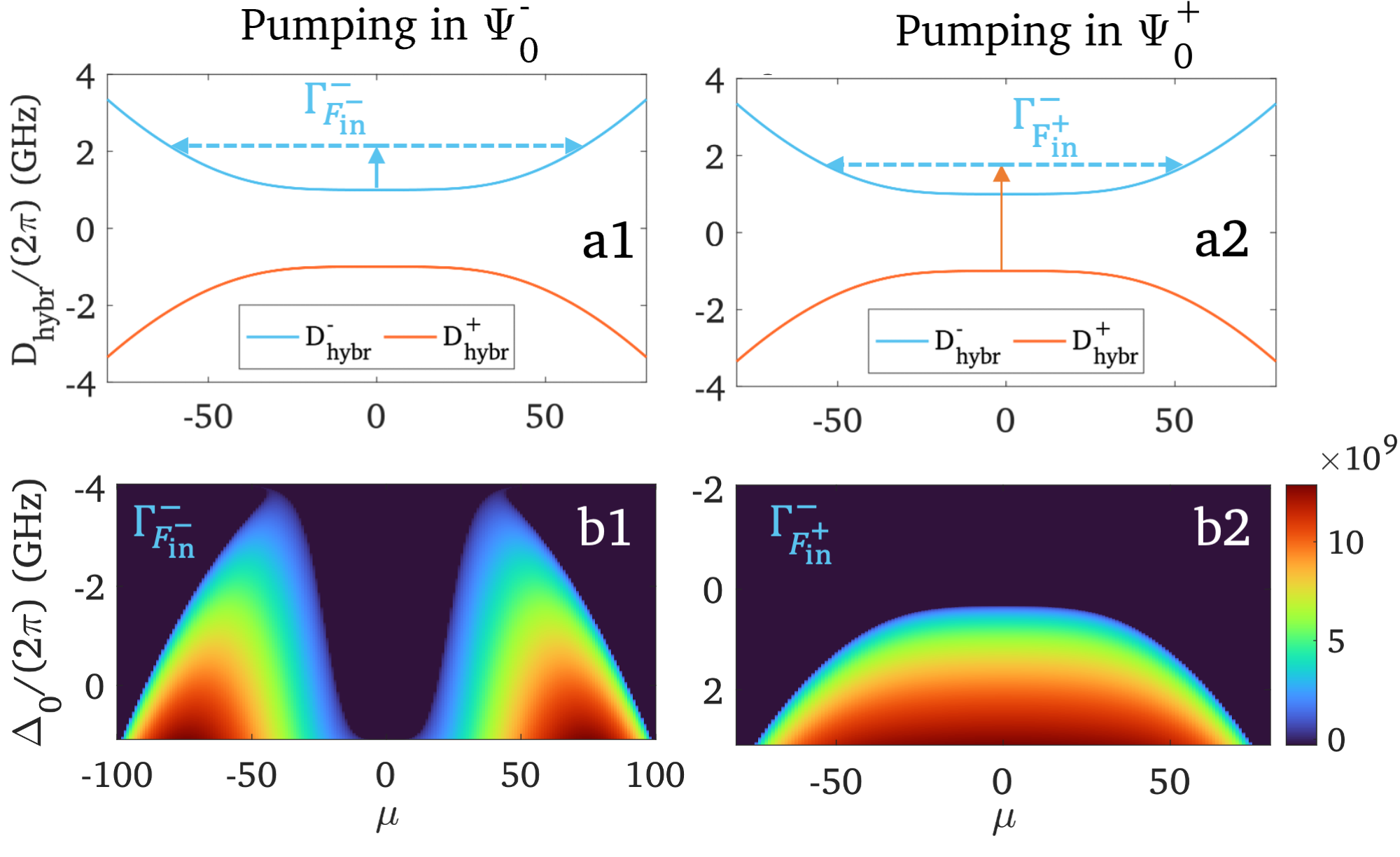}
    \caption{(a1,2) Integrated dispersion of the hybridized supermode families for the opposite sign dispersion of microresonators \(D_2^{\text{A}}/(2\pi)=1\) MHz, \(D_2^{\text{B}}/(2\pi)=-1\) MHz; arrows indicate the four-wave mixing processes for pumping in (a1) $\Psi_0^{-}$ and (a2) $\Psi_0^{+}$. MI gain of \(\Psi^-_\mu\) for pumping in (b1) $\Psi_0^{-}$ and (b2) $\Psi_0^{+}$ (\(\kappa = 100\) MHz, \(P_{\text{in}} = 0.3\) W, \(\Delta_{\text{res}}=0\)).}
    \label{fig:opposite_anom-norm}
\end{figure}

For example, we considered the case of opposite signs of dispersion with the same moduli \(D^{\text{A}}_2=-D^{\text{B}}_2\) and offset \(\Delta_{\text{res}}=0\), which ensures the accuracy of the analytical solution. The hybridized dispersion in this case \(D_{\text{hybr}}^\pm(\mu) =\mp\sqrt{(D_2^\text{A})^2\mu^4+4J^2}\) for supermodes \(\Psi^+_\mu\) is effectively negative for all \(\mu\) and for supermodes \(\Psi^-_\mu\) is effectively positive for all \(\mu\) (see Fig.\ref{fig:opposite_anom-norm}a). Moreover, unlike in identical microresonators, the sets of \(\Psi^-_\mu\) and \(\Psi^+_\mu\) modes do not lie on the same line, but are clearly separated. In this regard, modulation instability is possible only for the modes \(\Psi^-_\mu\) with positive effective dispersion. Here, for pumping into the supermode \(\Psi^-_0\), the gain lobes look like those for a single microresonator (see Fig.\ref{fig:opposite_anom-norm}b1), and when pumping into the supermode \(\Psi^+_0\), modulation instability occurs at large detunings, while the gain lobes follow the profile of the hybridized dispersion (see Fig.\ref{fig:opposite_anom-norm}b2). 

It is worth noting that we compared the results of a direct simulation of the system \eqref{General_MI} with the results of an analytical assessment and found them to be in exact agreement. Now, let us examine the extent to which the results of a direct system simulation might differ from the analytical predictions.

\begin{figure}
\centering
\includegraphics[width=\linewidth]{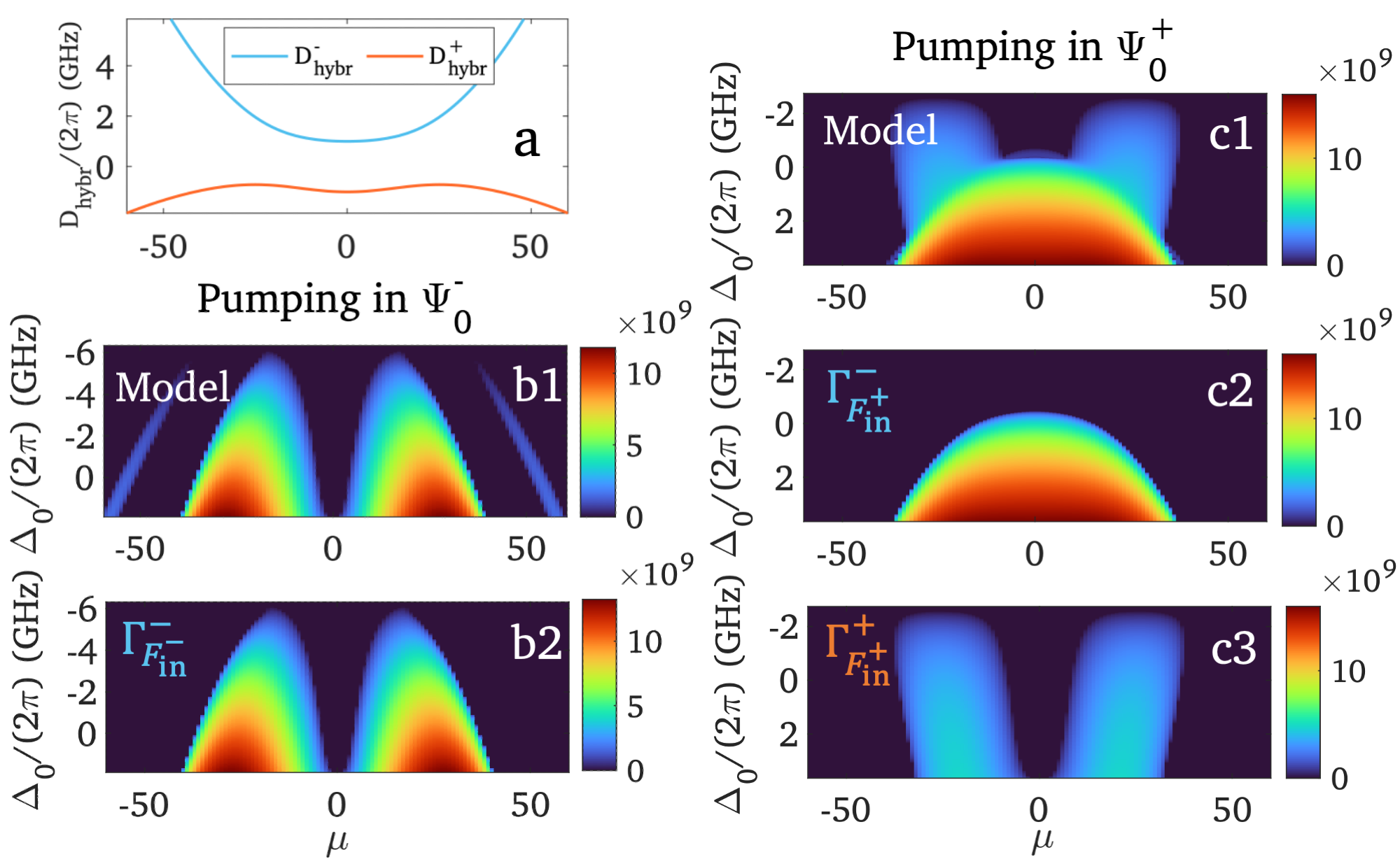}
    \caption{(a) Integrated dispersion of the hybridized supermode families for the dispersion of microresonators \(D_2^{\text{A}}/(2\pi)=5\) MHz, \(D_2^{\text{B}}/(2\pi)=-1\) MHz and absolute detuning \(\Delta_{\text{res}}/(2\pi)=-0.2\) GHz. Model MI gain  for pumping in (b1) $\Psi_0^{-}$ and (c1) $\Psi_0^{+}$. Analytical MI gain of $\Psi_\mu^{-}$ for pumping (b2) $\Psi_0^{-}$ and (c2) \(\Psi_0^{+}\). (c3) Analytical MI gain of $\Psi_\mu^{+}$ for pumping \(\Psi_0^{+}\) (\(\kappa = 100\) MHz, \(P_{\text{in}} = 0.45\) W).}
    \label{fig:anom-norm-0.2}
\end{figure}

As an example, we considered two resonators with dispersions of \(D_2^{\text{A}}/(2\pi)=5\) MHz and \(D_2^{\text{B}}/(2\pi)=-1\) MHz at different relative offsets \(\Delta_{\text{res}}\). For a small relative offset \(\Delta_{\text{res}}/(2\pi)=-0.2\) GHz, the analytical solution \eqref{eq:gain} provides a good prediction of the gain for the numerical solution of the system \eqref{General_MI}. For supermode \(\Psi^-_0\) pumping, we see a small amplitude gain for modes \(\mu=\pm[40,60]\) (see Fig. \ref{fig:anom-norm-0.2} b1), which is not predicted by the analytics. For supermode \(\Psi^+_0\) pumping, however, the simulation yields a superposition of the gains described in the analysis (see Fig. \ref{fig:anom-norm-0.2}c1). It is important to note that Figure \ref{fig:anom-norm-0.2}c1 shows the maximum increment, meaning that at the overlap points between Figures \ref{fig:anom-norm-0.2}c2 and \ref{fig:anom-norm-0.2}c3, the numerical simulation yields two solutions. 

\begin{figure}[htp!]
\centering
\includegraphics[width=\linewidth]{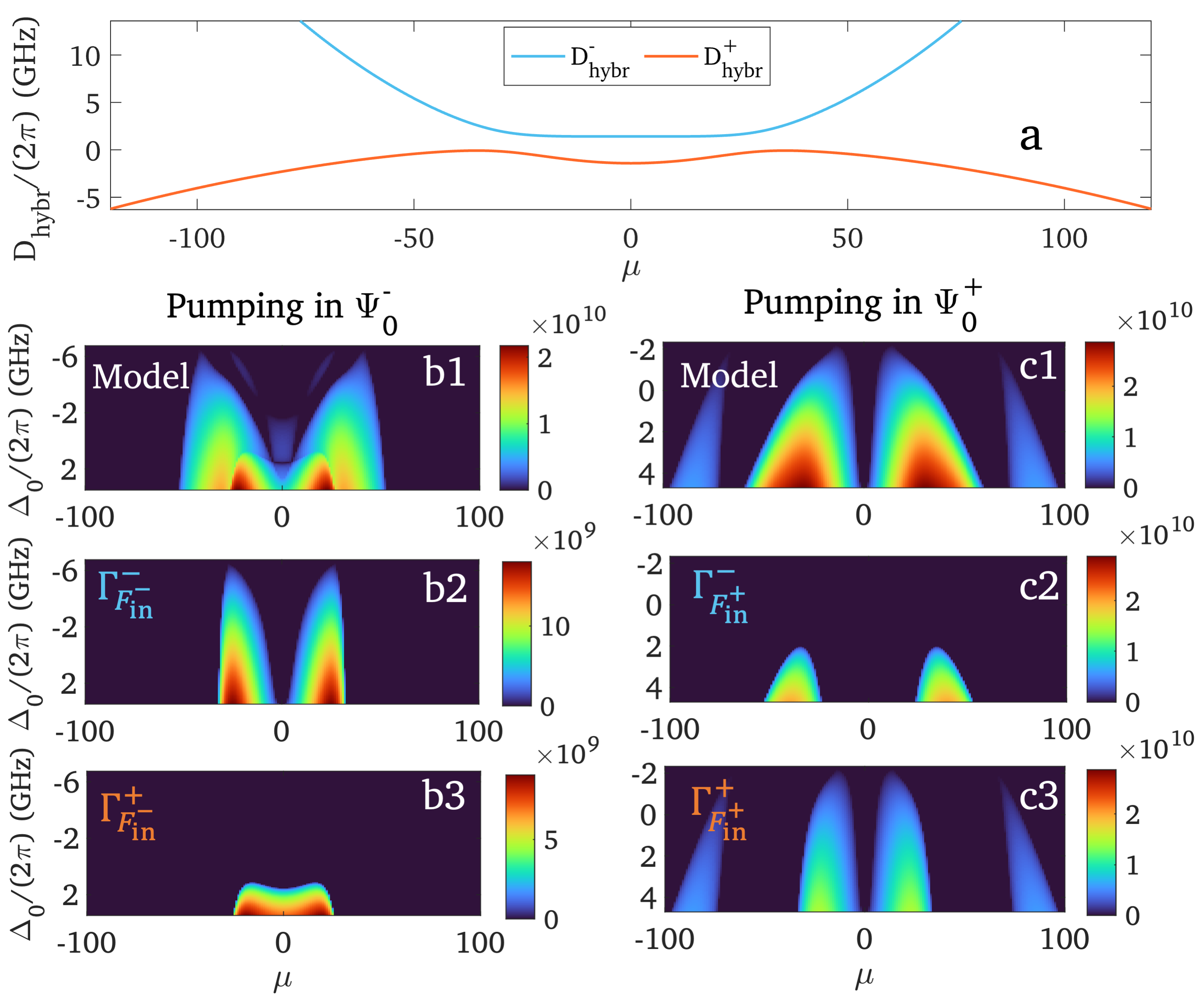}
    \caption{(a) Integrated dispersion of the hybridized supermode families for the dispersion of microresonators \(D_2^{\text{A}}/(2\pi)=5\) MHz, \(D_2^{\text{B}}/(2\pi)=-1\) MHz, and absolute detuning \(\Delta_{\text{res}}/(2\pi)=-2\) GHz. Model MI gain  for pumping in (b1) $\Psi_0^{-}$ and (c1) $\Psi_0^{+}$. Analytical MI gain of $\Psi_\mu^{-}$ for pumping (b2) $\Psi_0^{-}$ and (c2) \(\Psi_0^{+}\). Analytical MI gain of $\Psi_\mu^{+}$ for pumping (b3) \(\Psi_0^{-}\) and (c3) \(\Psi_0^{+}\) (\(\kappa = 100\) MHz, \(P_{\text{in}} = 0.45\) W).}
    \label{fig:anom-norm-2}
\end{figure}

With increasing relative detuning to \(\Delta_{\text{res}}/(2\pi)=-2\) GHz, a greater discrepancy arises between the model and the analytics, but the analytics still describe the main patterns of behavior of gain lobes (see Fig. \ref{fig:anom-norm-2}). 

\begin{figure}
\centering
\includegraphics[width=\linewidth]{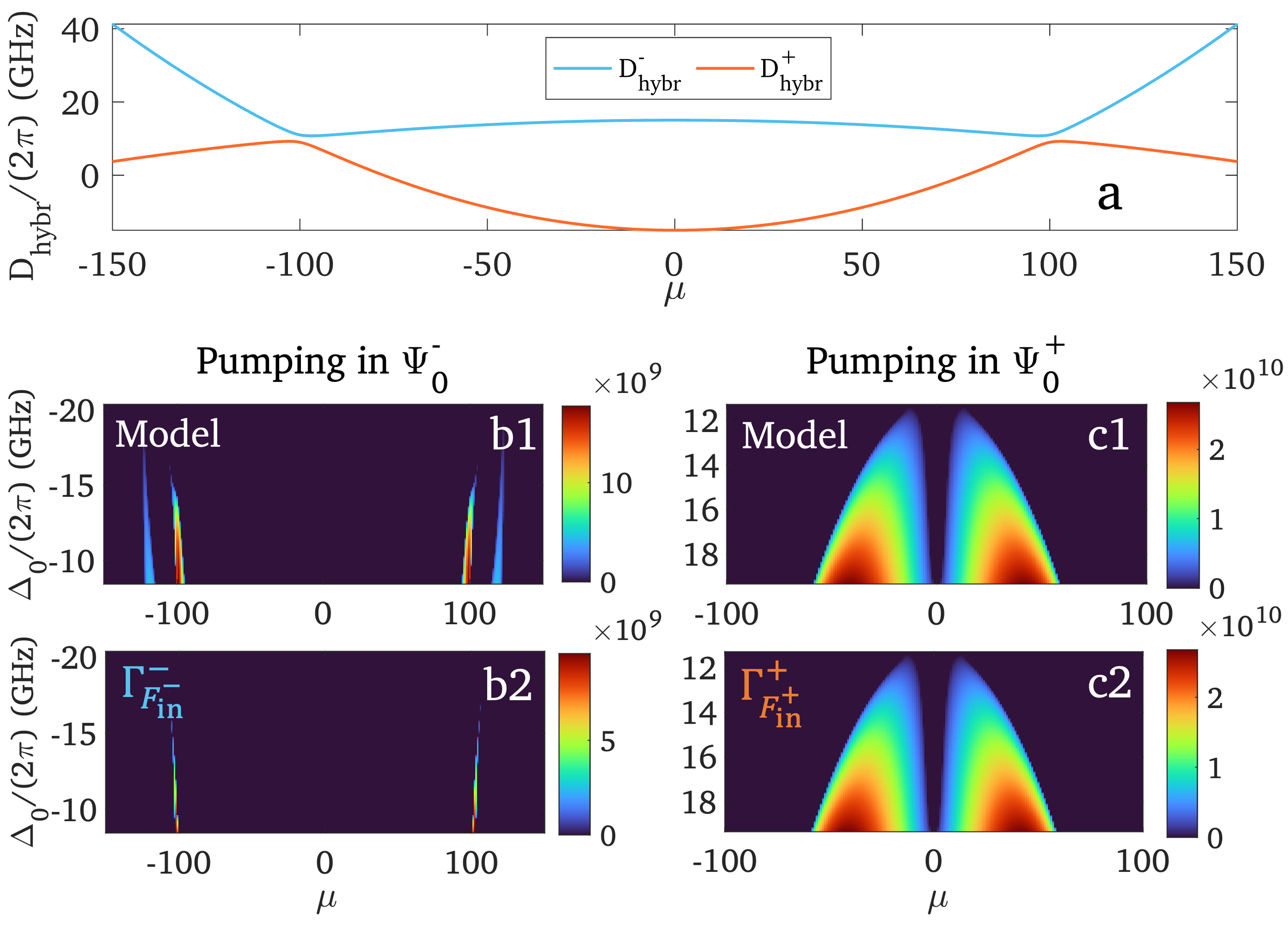}
    \caption{(a) Integrated dispersion of the hybridized supermode families for the dispersion of microresonators \(D_2^{\text{A}}/(2\pi)=5\) MHz, \(D_2^{\text{B}}/(2\pi)=-1\) MHz, and absolute detuning \(\Delta_{\text{res}}/(2\pi)=-30\) GHz. Model MI gain  for pumping in (b1) $\Psi_0^{-}$ and (c1) $\Psi_0^{+}$. (b2) Analytical MI gain of $\Psi_\mu^{-}$ for pumping  $\Psi_0^{-}$. (c2) Analytical MI gain of $\Psi_\mu^{+}$ for pumping \(\Psi_0^{+}\) (\(\kappa = 100\) MHz, \(P_{\text{in}} = 0.45\) W).}
    \label{fig:anom-norm-30}
\end{figure}

For the limiting case of large detuning \(\Delta_{\text{res}}/(2\pi)=-30\) GHz, this leads to the fact that, in the region of modes for which modulation instability is possible, the set of supermodes \(\Psi^-_\mu\) effectively has negative dispersion, while the set of supermodes \(\Psi^+_\mu\) effectively has positive dispersion (see Fig. \ref{fig:anom-norm-30}a). Therefore, the analysis again describes the gain lobes of the supermodes \(\Psi^+_\mu\) well (see Fig. \ref{fig:anom-norm-30} c1,c2), while for the supermodes \(\Psi^-_\mu\), very narrow gains appear at the points where the effective dispersion changes sign near \(\mu\sim\pm100\) (see Fig. \ref{fig:anom-norm-30} b1,b2).

It is important to note that the shape of the gain lobes is primarily determined by the hybridized dispersion profile. The contribution of different losses to the change in dispersion shape is insignificant, so for \(\kappa^{\text{A}}\neq\kappa^{\text{B}}\), the lobe shape changes only slightly, but the absolute value of the gain changes due to hybridized losses.

\section{Conclusions}
Coupled optical resonator waveguide systems, also known as CROWs, are complex optical devices with many applications in integrated photonics thanks to their spectral selectivity and nonlinear response. Successful deployment of delay lines and division multiplexing filters demonstrated tight control over coupled-microresonator fabrication and paved the way for practical use of nonlinear wavelength conversion \cite{morichetti2012first}. Linear approaches for the analysis of CROW systems have been proposed previously \cite{capmany1990new}. Here, to facilitate progress in numerical modeling and experimental studies of the nonlinear regime of coupled Kerr microresonators \cite{tikan2021emergent,komagata2021dissipative,pidgayko2023voltage,kolesnikova2026hybridization}, we provide a theoretical basis for modulation instability in anomalous-anomalous, normal-normal, and anomalous-normal microrings.

The coupled ring system has two pump resonances, from which the pumped photons can be converted via intra- and inter-band processes, yielding four MI gain characteristics $\Gamma^\pm_{F^\pm_{\text{in}}}$, which we have investigated in this work. Thanks to the near-vanishing antidiagonal elements of the MI eigenvalue problem, we provide analytical expressions for the gain curves that remain valid until waveguide geometries differ markedly, which is mostly relevant for the anomalous-normal configuration. In the latter case, our theory still provides fundamental understanding of the four-wave mixing process. At the same time, the exact shape of the MI gain can be extracted numerically by solving the eigenvalue problem for system (\ref{General_MI}). 

Central to these results is the generalized supermode basis, which is the natural basis for coupled microresonators. In the basis of individual resonator modes, the stationary pump field is distributed between the rings in a way that depends on the pump amplitudes, their relative phase, and the detuning, and this distribution enters every coefficient of the linearized problem. In the supermode basis, by contrast, the linear part is diagonal, the pump detuning alone selects which supermode is excited, and the ratio of the input amplitudes only sets the pumping efficiency. The problem then takes a single-resonator-like form with hybridized dispersion $D^{\pm}_{\text{hybr}}(\mu)$, losses $\kappa^{\pm}_{\text{hybr}}(\mu)$, and effective nonlinearity $G_0^{\pm}$, so that the MI threshold and gain lobes can be read off directly from the hybridized dispersion profile. 

The results obtained enable identification of the modulation instability region, together with the bistability area in the two-dimensional parameter space $(\Delta_0, P_{\text{in}})$, as shown in Figure~\ref{fig:mi_region}, which is of utmost importance for managing the nonlinear operating regime in coupled Kerr resonators, including OPO and DKS generation. For example, our analysis shows that pumping the \(\Psi_0^-\) supermode of identical anomalous-anomalous microresonators results in the almost simultaneous parametric generation of both supermodes, which is in line with recent experimental and numerical studies showing coexistence of coherent structures and dispersive waves from different supermodes \cite{tikan2021emergent,komagata2021dissipative}. A different situation arises in the anomalous-normal configuration: at a large resonance offset $\Delta_{\text{res}}/(2\pi)=-30$~GHz (Fig.~\ref{fig:anom-norm-30}), MI of the $\Psi^-_\mu$ supermodes develops only in very narrow gain bands located where the effective dispersion changes sign, a case which was recently investigated numerically and experimentally in \cite{pidgayko2023voltage}. Such spectrally selective parametric gain, confined to a single pair of sidebands far from the pump, may be beneficial for the selective generation of squeezed light \cite{PhysRevApplied.3.044005}.

Since the supermode formalism relies only on diagonalizing the linear coupled-mode problem, our model can be straightforwardly generalized to the case of nested microresonators with different free spectral ranges (FSRs, \(D_1^{\text{A}}\neq D_1^{\text{B}}\)) by replacing \(D_2^{\text{A,B}}\mu^2/2\rightarrow D_1^{\text{A,B}}\mu+D_2^{\text{A,B}}\mu^2/2\). Generalization to the case of coupled microresonators with different FSRs and a coupling section, see, e.g., recent experimental work on such systems \cite{yuan2023soliton}, is possible; however, it requires numerically determining the supermode eigenfrequencies and their spatial distributions across the microresonators. Finally, increasing the complexity of the system by adding more rings to the problem setting, up to and including CROW chains and lattices, is numerically feasible, yielding complex, multi-dimensional MI gain characteristics \cite{tusnin2023nonlinear}.

\begin{acknowledgments}
This work was supported by the Russian Science Foundation project N\textsuperscript{\underline{\scriptsize o}} 25-72-31023.
\end{acknowledgments}

\onecolumngrid
\section{Appendix}

In the case of different losses \(\kappa^{\text{A}}\neq\kappa^{\text{B}}\) coefficients \(C_\mu^\pm\) are complex. For pumping into into supermode \(\Psi^-_0\), positive increment for supermodes \(\Psi^\pm_\mu\) is defined by
\begin{equation}
    \lambda^{\pm}_{F^-_{\text{in}}}(\mu)=-\cfrac{\kappa_{\text{hybr}}^\pm(\mu)}{2}-2\text{Im}\left(\Tilde{G}_{F^-_{\text{in}}}^{\pm}(\mu)\right)|\Psi^-_0|^2+\sqrt{\left|G_{F^-_{\text{in}}}^{\pm}(\mu)\right|^2|\Psi^-_0|^4-\left(\Delta_0+\cfrac{D_{\text{hybr}}^\pm(\mu)}{2}-2\text{Re}\left(\Tilde{G}_{F^-_{\text{in}}}^{\pm}(\mu)\right)|\Psi^-_0|^2\right)^2}, 
\end{equation} 
where $G^\pm_{F^-_{\text{in}}}(\mu)=(1+|C_0^-|^2)\,\cfrac{g^A(C_0^+)^2\,C_\mu^{\mp*}-g^B C_\mu^\pm}{\big(C_\mu^{\mp*}-C_\mu^{\pm*}\big)\,(C_0^--C_0^+)^2}$, \(\Tilde{G}_{F^-_{\text{in}}}^\pm(\mu) = \cfrac{g^{\text{A}}|C_0^{+}|^2(1+|C_0^-|^2)C_\mu^\mp-g^{\text{B}}(1+|C^-_0|^2)C_\mu^\pm}{(C^\mp_\mu-C^\pm_\mu)|C_0^--C_0^+|^2}\). 

For pumping into supermode \(\Psi^+_0\), an analogous positive increment for supermodes \(\Psi^\pm_\mu\) reads as
\begin{equation}
    \lambda^{\pm}_{F^+_{\text{in}}}(\mu)=-\cfrac{\kappa_{\text{hybr}}^\pm(\mu)}{2}-2\text{Im}\left(\Tilde{G}_{F^+_{\text{in}}}^{\pm}(\mu)\right)|\Psi^+_0|^2+\sqrt{\left|G_{F^+_{\text{in}}}^{\pm}(\mu)\right|^2|\Psi^+_0|^4-\left(\Delta_0+\cfrac{D_{\text{hybr}}^\pm(\mu)}{2}-2\text{Re}\left(\Tilde{G}_{F^+_{\text{in}}}^{\pm}(\mu)\right)|\Psi^+_0|^2\right)^2}, 
\end{equation} 
where \(G^\pm_{F^+_{\text{in}}}(\mu)=(1+|C_0^+|^2)\,\cfrac{g^{\text{A}}(C_0^-)^2\,C_\mu^{\mp*}-g^{\text{B}}C_\mu^\pm}{\big(C_\mu^{\mp*}-C_\mu^{\pm*}\big)\,(C_0^--C_0^+)^2}\),
\(\Tilde{G}_{F^+_{\text{in}}}^\pm(\mu) = \cfrac{g^{\text{A}}|C_0^{-}|^2(1+|C_0^+|^2)C_\mu^\mp-g^{\text{B}}(1+|C^+_0|^2)C_\mu^\pm}{(C^\mp_\mu-C^\pm_\mu)|C_0^--C_0^+|^2}\).

In the case of equal losses, \(\Tilde{G}_{F^\pm_{\text{in}}}^\pm(\mu)\equiv G_{F^\pm_{\text{in}}}^\pm(\mu)\) from main text and \(\text{Im}\left\{G_{F^\pm_{\text{in}}}^\pm(\mu)\right\}\equiv0\) and \(\text{Im}\left\{\Tilde{G}_{F^\pm_{\text{in}}}^\pm(\mu)\right\}\equiv0\).

%

\end{document}